# Frank Ramsey's Contributions to Probability Theory and Legal Theory

(Review of Cheryl Misak, *Frank Ramsey: A Sheer Excess of Powers*, Oxford U Press, 2020)

F. E. Guerra-Pujol
*University of Central Florida*
fegp@ucf.edu

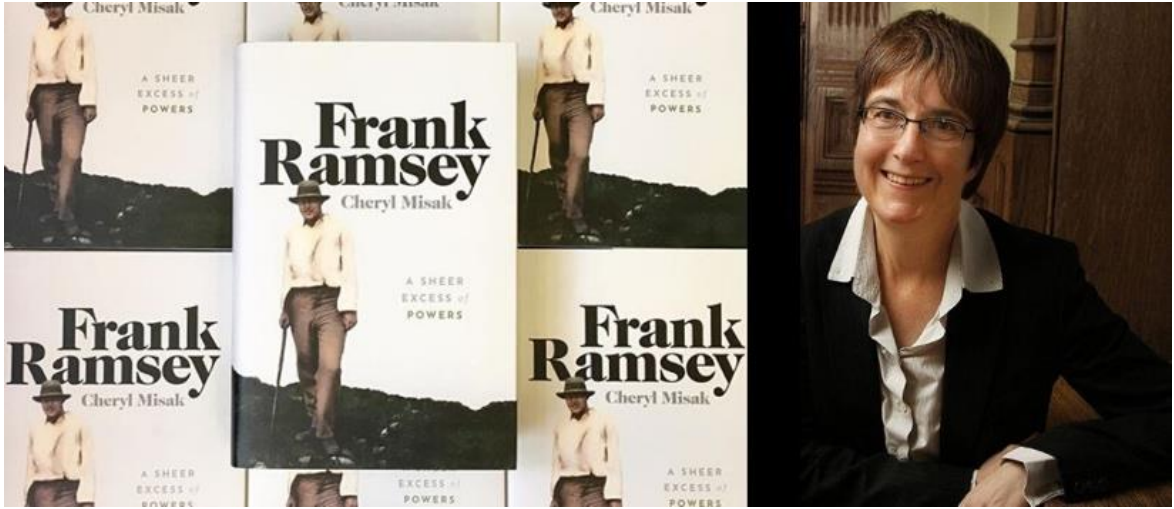

| | |
|---|---|
| Abstract | The English polymath Frank Ramsey was one of the first scholars to paint a subjective picture of probability, but how and when did he make this discovery? Among other things, Cheryl Misak's beautiful biography of Ramsey explores this remarkable terrain, which we review in part one of this paper. In addition, this review will explore the significance of Ramsey's contributions to law and legal theory. Building on my previous work, I will present a Ramsian method of judicial voting as well as a simple Ramsian model of appellate litigation in part two of this paper. |
| Keywords | Thomas Bayes, Cheryl Misak, Frank Ramsey, degrees of belief, litigation, subjective probability |
| JEL Codes | B29, B31, C11, K41 |
| Date | July 13, 2020 |
| Word Count | 5,453 |

Contents

*"A query sent out to Twitter, asking for innovations named after Ramsey, produced an astonishing nineteen items."*
--Cheryl Misak (2020b)

*Introduction*

Among other things, Frank Ramsey (b. 1903, d. 1930) was one of the first scholars, along with Bruno de Finetti (1974), to formalize subjective probability. In brief, the subjective view of probability can be expressed in terms of "degrees of belief." By way of example, the probability of x outcome, where x a unique event such as Disney World reopening by a certain date or the U.S. Supreme Court voting to overrule *Roe v. Wade*, need not be an objective value but rather can consist of an individual's personal or subjective judgment about whether x outcome is likely to occur.

But how did Ramsey discover this revolutionary insight in the first place--the idea that probability can consist of a subjective or personal value? Cheryl Misak's new biography of Frank Ramsey (Misak, 2020a), which is subtitled "A Sheer Excess of Powers," explores this terrain as well as Ramsey's many other scholarly contributions. Misak's book is divided into three broad temporal sections: (1) "Boyhood," which consists of three chapters devoted to the years 1903 to 1920, i.e. from the year of Ramsey's birth up to his arrival at Cambridge University; (2) "The Cambridge Man," which contains seven chapters that describe Ramsey's undergraduate years at Trinity College as well as his six-month sojourn in Vienna in 1924; and lastly (3) "An Astonishing Half Decade," which contains nine chapters and covers the last five years of Ramsey's short but productive life.

I will likewise divide my review of Misak's biography in part one of this paper into three parts, with each part of my review corresponding to one of the sections of Misak's book. (Also, although Ramsey made significant contributions to a wide variety of fields, including economics, mathematics, and philosophy, I will focus the rest of my review on Ramsey's contributions to probability theory.) Next, in part two of this paper, I will explain how Ramsey's theory of subjective probability sheds new light on two areas of law: litigation and judicial voting.

PART ONE: REVIEW OF MISAK

*Ramsey's boyhood*

Although there is no direct evidence that Ramsey was exposed to the rigors of probability theory by his parents Arthur and Agnes Ramsey or during his formal education at Winchester College (a demanding English boarding school for boys), two details from Ramsey's boyhood, as recounted in Part I of Misak's beautiful book, stood out for me the most. One was the young Ramsey's principled opposition to the brutal system of bullying and hazing at his boarding school. Misak summarizes this savage system on pp. 30-31 of her book. (Unless otherwise noted, all page references in this review refer to Misak's 2020 biography of Ramsey.) Here is just one telling excerpt:

> Each junior was the personal servant of an older student and had to "fag" or "sweat" for him. That meant cleaning the buttons and boots of his Officers' Training Corps uniforms as well as his muddy cricket boots so they gleaned white again, as well as countless other tasks. The juniors [also] had to make the prefects' tea, or afternoon meal, and wash up after ….

Ramsey detested these regular hazing rituals, and towards the end of his tenure at boarding school, he wrote in his diary that he had "[d]ecided to give up sweating juniors." According to Misak (p. 49), Ramsey even made a bargain with the younger boy assigned to him that he would not be required to do any chores at all for him; in return, the boy was to pay it forward to his assigned junior. In other words, in this episode we see the young Ramsey was a man of principle.

The other aspect of Ramsey's boyhood that caught my attention was his voracious reading habits, which Misak describes on pp. 48-49 of her book. Even at such an early age, Ramsey was a boy who loved the world of ideas, for in addition to his regular coursework at his boarding school, Ramsey devoured dozens of advanced works from a wide variety of fields. Among the many extracurricular books the young Ramsey is reported to have read are David Hume's *Treatise on Human Nature*, Bertrand Russell's *Problems of Philosophy*, and G.E. Moore's *Ethics*. If one were to create a syllabus with the goal of imparting a well-rounded and liberal arts education, one would be hard-pressed to assemble a better collection of classic works.

*Cambridge University*

The next chapter in Ramsey's intellectual life would take place at Cambridge University, where he would, among other things, study John Maynard Keynes's *Treatise on Probability* and challenge Keynes's approach to probability theory. Part II of Cheryl Misak's intellectual biography of Frank Ramsey is thus devoted to the young Ramsey's undergraduate years at Cambridge. If there is a common or overarching theme during these formative years in Ramsey's intellectual life (1920 to 1924), it is his willingness to challenge the most powerful and sacrosanct ideas of such great and legendary scholars and philosophers as G.E. Moore, Bertrand Russell, and Ludwig Wittgenstein. Here, I will limit myself to just one such momentous undergraduate episode: Ramsey's early critique of Keynes's theory of probability.

To appreciate Ramsey's first foray into probability theory, I must first provide some relevant background. The great John Maynard Keynes had published his *Treatise on Probability* in 1921, and in a review of Keynes's work, none other than Bertrand Russell had called Keynes's *Treatise* "the most important work on probability that has appeared for a very long time," adding that the "book as a whole is one which it is impossible to praise too highly." (See Russell, 1948 [1922], p. 152.) Why was Keynes's work so highly praised? Because Keynes had developed a new way of looking at probability, one which allowed for the possibility of probabilistic truth. For Keynes, probability consisted of an objective or logical relation between evidence and hypothesis, or in the words of Misak (2020, p. 113, emphasis added), a relation "between any set of premises and a conclusion in virtue of which, *if we know the first*, we will be warranted in in accepting the second with some particular degree of belief."

Ramsey, however, immediately identified two blind spots in Keynes's conception of probability. (See Misak, 2020, pp. 114-115. For the record, Ramsey published his review of Keynes's *Treatise on Probability* in the January 1922 issue of *Cambridge Magazine*, while he was still an undergraduate!) One was Keynes's admission that not all probabilities are numerical or measurable, especially when the truth values of our underlying premises are in dispute. In that case, when we have no idea whether our premises are true or not, Keynes's approach does not allow us to measure the probabilities of our conclusions. For Ramsey, by contrast, all probabilities should be measurable. But the other (more deeper) problem with Keynes's theory was the "objective" nature of his view of probability–the idea that all statements or propositions stand in logical relation to each other. Ramsey denied the existence of these logical relations altogether. Far from being an "objective relation," the strength or weakness of the relationship between two propositions also depended on psychological factors: on one's personal experiences and subjective beliefs. In a word, probability was based on *experience*, not logic. For students of the common law (myself included), Ramsey's approach to probability may sound familiar, or in the immortal words of the great Oliver Wendell Holmes (1991, p. 1): "The life of the law has not been logic: it has been experience."

Yet, as the saying goes, *it takes a theory to beat a theory*, and at this stage of his promising career the young Ramsey had yet to develop his own full-fledged theory of probability. Ramsey would finally get around to doing so in the last half decade of his short life, but before proceeding, let's take a short detour to recount two important personal episodes in Ramsey's life: his six-month sojourn in Vienna (see chapter 7 of Misak), and his secret love affair with Lettice Baker (pp. 205-208), who deserves a biography of her own.

In brief, upon the completion of his undergraduate studies Ramsey had decided to spend an extended period of time in Vienna in order to undergo psychoanalysis. (It was during this time that Ramsey received the news of his appointment to a lectureship at King's College. See Misak, pp. 178-181.) According to Misak (p. 161), at that time "taking the cure in Vienna" was a common pastime for many young Cambridge academics. In addition, Frank Ramsey took full advantage of all that the former imperial capital had to offer, including deep discussions with members of the legendary "Vienna Circle" of anti-metaphysical philosophers, cultured nights at the world-famous Opera, and even some sordid sexual escapades with a Viennese prostitute.

More importantly, within a month of his return to England in the fall of 1924 to begin his teaching duties, Ramsey met Lettice Baker at a Moral Sciences Club meeting at Trinity College. (It was at this meeting that G. E. Moore read his now famous paper "A defence of common sense." See Misak, p. 205.) Shortly thereafter, Ramsey asked her out to tea, and they quickly fell in love. Among other things, one of the things that struck me the most about the Ramsey-Baker love affair was how they had to keep their amorous relationship a secret (pp. 207-208) until they formally got married in September of 1925. Be that as it may, Ramsey's greatest contributions to the world of ideas, including his theory of subjective probability, were right around the corner.

*Ramsey's swan song*

The third and last part of Cheryl Misak's beautiful biography of Frank Ramsey ("An Astonishing Half Decade") covers the last five years of Ramsey's life. During the last half decade of his short but productive life, Ramsey made major contributions to a wide variety of fields, including economics, mathematics, and philosophy, but I shall focus here on his contributions to probability theory, for it was during this time that Ramsey developed his own full-fledged theory of "subjective" or psychological probability.

Ramsey developed his new approach to chance in a paper titled "Truth and Probability," which he presented for the first time at a meeting of the Moral Sciences Club in November of 1926. (See Misak, 2020, p. 263. Ramsey's influential 1926 paper was eventually published posthumously in 1931.) In this remarkable paper, Ramsey sketched out an entirely new and revolutionary way of looking at probability. We can summarize Ramsey's picture of probability in ten words: *probabilities are beliefs and beliefs, in turn, are metaphorical bets*. For example, in Ramsey's own words (quoted in Misak, p. 268), "Whenever we go to the station we are betting that a train will really run, and if we had not a sufficient degree of belief in this [outcome] we should decline this bet and stay at home."

On this subjective view of probability, we can measure the strength of a person's personal beliefs in betting terms, or again in Ramsey's own words (Misak, p. 271), a "probability of 1/3 is clearly related to the kind of belief [that] would lead to a bet of 2 to 1." In addition, Ramsey showed how one's bets–i.e. one's subjective or personal probabilities–should obey the formal axioms of probability theory. (As as aside, Misak's biography includes a separate summary by Nils-Eric Sahlin of the technical details of Ramsey's subjective or betting approach to probability. See Misak, pp. 272-273. See also Hájek, 2019.) It is hard to understate the importance of Ramsey's "betting paradigm" of probability. Before concluding my review, then, allow me to call a time out to explain why Ramsey's subjective or psychological theory of probability is so "significant" (pun intended, for my frequentist friends).

First and foremost, Ramsey's betting paradigm fills a huge gap left open by standard probability theory (i.e. frequency theory), for as Misak herself correctly notes (p. 266), frequentist methods can neither "provide an account of partial belief, nor an account of how an individual should make one-off decisions." This blind spot is so enormous and so well-known by now that I won't belabor it here. Instead, it suffices to say that Ramsey single-handedly delivered a serious blow to frequency theory when he developed his subjective theory of probability. Secondly, Ramsey's betting paradigm fills another huge blind spot in probability theory. Before Ramsey, the conventional wisdom so to speak was that some probabilities (especially personal probabilities) were not measureable in any rigorous way. After Ramsey, by contrast, we are now fully able to express any person's partial beliefs, even his subjective ones, using numerical values. How? By converting one's beliefs into betting odds. (As an added bonus, Ramsey's intellectual framework even enables us to determine whether our subjective beliefs are rational are or not, via now-standard Dutch book arguments. See, e.g., Vineberg, 2016.)

In addition, Ramsey's betting paradigm also provides the intellectual foundations for modern prediction markets (see, e.g., Wolfers & Zitzewitz, 2004), one of the most promising and exciting "mechanism designs" or market innovations of our time (see Arrow, et al., 2008). Perhaps we should rename prediction markets "Ramsey markets" in honor of Frank Ramsey. In any case, I have blogged about prediction markets many times before, so I won't belabor this third point here. And last, but not least, Ramsey's subjective approach to probability can be used to improve collective decision making via a method often referred to as "range voting" (see Smith, 2000) or "utilitarian voting" (see Hillinger, 2005) or to improve legal adjudication as well via similar methods I have christened "Bayesian verdicts" (see Guerra-Pujol, 2015) and "Bayesian judging" (see Guerra-Pujol, 2019). Again, perhaps we should rename these alternative methods of voting "Ramsian voting" or "Ramsey voting" in honor of Frank Ramsey. Regardless of the label we choose for these innovations, I can't overstate enough how important and exciting Ramsey's subjective approach is.

In the remainder of this paper (part two), I will turn to the world of legal theory and explain how Ramsey's insights about subjective probability also illuminate two areas of law: litigation and judicial voting.

PART TWO: FRANK RAMSEY'S CONTRIBUTIONS TO LEGAL THEORY

*Ramsian model of litigation*

Building on my previous work (Guerra-Pujol, 2011), I shall present a simple Ramsian model of litigation. In brief, Ramsey's methods allow us to estimate the probability distribution of the outcome of appellate cases, even with very little information, e.g. a small sample of previously-decided cases.

To illustrate Ramsey's methods, I shall use a small sample of federal appellate cases collected in my previous work (Guerra-Pujol, 2011), i.e. the set of all reported cases in Volume 287 of the Federal Reporter, excluding criminal cases. [In future work, I will sample the set of all cases decided by the United States Supreme Court during the 2019-2020 Term.] In all, our sample of appellate cases contained 54 non-criminal "government cases" and 103 "non-government cases" that were initially decided at various stages during the initial litigation: pre-trial, trial, and post-trial. [See Tables 2AA and 2D in Guerra-Pujol, 2011, pp. 58-59. The term "trial" includes non-jury adjudications, such as injunction hearings and other equity cases. We are excluding criminal cases from our sample since plea bargains are not reported in the Federal Reporter.]

From a probabilistic perspective, this sample of appellate cases in Guerra-Pujol (2011) is like a "legal urn" or black box containing a set of 154 federal cases (events) instead of 154 balls. Based on our modest sample, our legal urn contains the following mix of cases (cf Schlaifer, 1959, p. 330-331):

- Just under $\frac{1}{2}$ (i.e. 71/148 or .48) of the cases were decided or disposed of during the **pre-trial stage**; of these cases, about $\frac{1}{5}$ (i.e. 15/71 or .21) were non-criminal government cases; about $\frac{4}{5}$ (i.e. 56/71 or .79) were non-government cases.

- A little over $\frac{1}{2}$ (i.e. 86/148 or .52) of the cases were decided at **trial or post-trial**; of these cases, over $\frac{2}{5}$ (39/86 of .45) were non-criminal government cases; a little less than $\frac{3}{5}$ (47/86 or .55) were non-government cases.

We can visualize these priors in Figure 1 below as follows:

---

[FIGURE 1 – PRIOR PROBABILITIES]

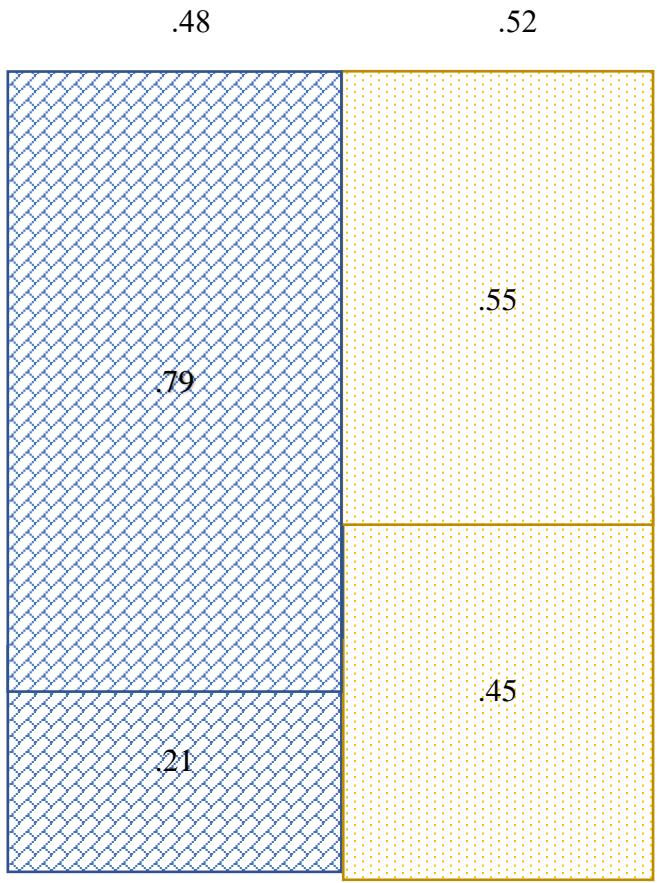

---

The values above the box represent the prior probabilities [or unconditional relative frequencies] that a federal case will be decided at the pre-trial stage or trial/post-trial stage, while the four values inside the box represent the conditional probabilities [or conditional relative frequencies] that any given case is either a non-criminal government case or a non-government case. [Although our sample consists of appellate cases, we carefully determined at which stage the case was decided before it went up to appeal. Also, we have equalized the first set priors (above the box) for mathematical simplicity.] This prior distribution is depicted in Table 1 below as follows:

---

[TABLE 1 – PRIOR PROBABILITIES IN TABULAR FORM]

| Event of interest | Column A: Prior Probability of event* | Column B: Probability the case is a (non-criminal) government case given the event |
|---|---|---|
| Pre-trial disposition | .48 | .21 |
| Disposition at trial or post-trial | .52 | .45 |
|  | .48 + .52 = 1.0 |  |

[Note: Column A is essential to avoid the base rate fallacy (see generally Bar-Hillel, 1980).]

---

Next, suppose a new case is now pending before a federal court (or suppose we draw any federal case at random), and further suppose that the only additional information we have is the identity of the parties, i.e. we have no idea whether the case was decided during the pre-trial stage or during the trial/post-trial stage. What is the probability that a case decided during the pre-trial stage is a government case? The solution to this problem requires us to revise or update our priors--using the two sets of prior probabilities shown in Table 1--as follows. First, we use the multiplication rule to compute the joint probabilities of "pre-trial decision" and "non-criminal government case" (i.e. .48 times .21 = .1008) and "trial/post-trial decision" and "non-criminal government case" (i.e. .52 times .45 = .2340). These joint probabilities are shown in column C of Table 2 below. We can then use the addition rule to compute the marginal probability of that a case will be a non-criminal government case. Lastly, we apply the definition of conditional probability to compute the revised or posterior probabilities in column D of Table 2:

---

[TABLE 2 – POSTERIOR PROBABILITIES IN TABULAR FORM]

| Event of interest | Column C: Joint probability of event and (non-criminal) government case | Column D: Probability of event given the case is a (non-criminal) government case |
|---|---|---|
| Pre-trial disposition | .1008 ≅ .1 | .1/.3 = .33 |
| Disposition at trial or post-trial | .2340 ≅ .2 | .2/.3 = .66 |
|  | .1 + .2 = .3 (as per addition rule) | .33 + .66 ≅ 1 |

---

Table 2 can also be depicted in visual form. The information in Column C of Table 2 is represented in visual form in Figure 2 below:

---

[FIGURE 2 – JOINT PROBABILITIES]

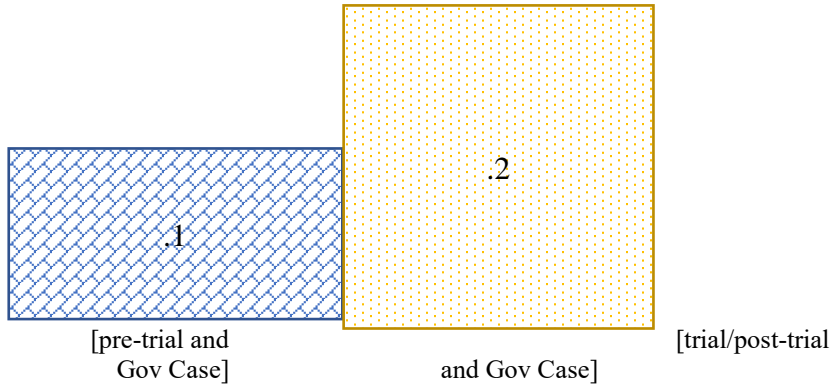

[pre-trial and Gov Case]   .1            .2   [trial/post-trial and Gov Case]

---

Figure 2 reproduces only that part of Figure 1 that corresponds to the event "non criminal government case" decided at either the pre-trial stage or trial/post-trial stage case. Next, we create Figure 3 by enlarging Figure 2 in such a way that its total area becomes 1. How do we do this? First, we first calculate the original area of each part of Figure 2 (the joint probabilities) from the data in Figure 1 and then divide the area of each part of Figure 2 by the total of the areas in Figure 2 [.3; the sum of .1 and .2] to raise the revised area to 1. (Cf. Schlaifer, 1959, p. 332.) The information in Column D of Table 2 is thus represented in visual form in Figure 3 below:

---

[FIGURE 3 – POSTERIOR PROBABILITIES]

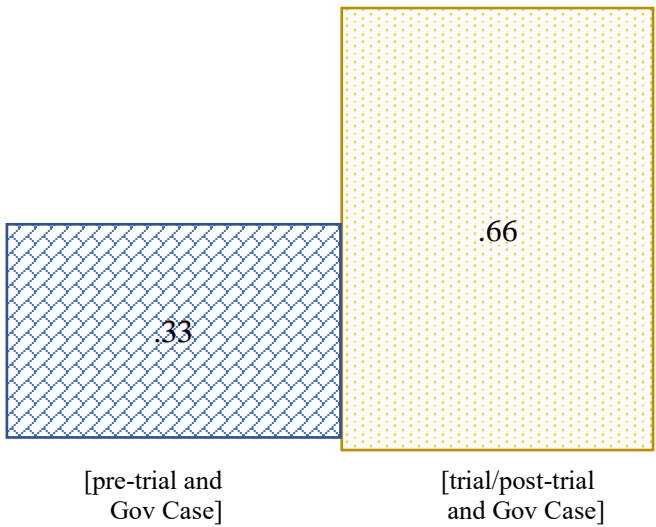

[pre-trial and Gov Case]   .33            .66   [trial/post-trial and Gov Case]

---

That is to say, given our base rates--i.e. the ratio of pre-trial cases to trial/post-trial cases--the probability that a case is a government case given that it was decided or disposed of during the trial/post-trial stage (instead of the pre-trial stage) is high. In other words, the government not only has a high probability of winning its cases (Guerra-Pujol, 2011). There is also a high probability that it will win its cases during the trial or post-trial stage.

To sum up this Ramsian model of appellate litigation, the advantage of the Ramsey's approach is that it allows us to make the best use of a minimum amount of information and experience. Of course, it is no substitute for experience, since it, by itself, cannot determine the probability of any event. [Cf. Schlaifer, 1959, p. 333.] Instead, our Ramsian formulation allows us to make more effective use of our experience by assigning probabilities to those events of which we have some experience or prior knowledge, rather than to events that may in reality determine litigation outcomes but with which we have no direct knowledge or experience.

*Ramsian model of judicial voting*

As we have seen, Frank Ramsey (along with de Finetti) was the first theorist to propose a subjective definition of probability, now referred to as "Bayesian probability." (See, e.g., Cox, 1946.) According to this Bayesian or *subjective* view of probability, probabilities are not an objective property of the real world. Instead, probabilities are simply the subjective expression of one's personal view of the world. In other words, the probability of a particular proposition being true is just a particular individual's degree of belief in the truth of that proposition. On this subjective view of probability, even if two people's judgments about the probability of a proposition are vastly different at time $t_1$, after evidence for (or against) the statement/hypothesis is introduced at time $t_2$, rational individuals should then revise their initial degrees of beliefs. Moreover, according to the subjective view, their degrees of belief will tend to converge to the same probability as more and more evidence comes in. In short, isn't this subjective convergence toward truth a good description of how common law judges decide cases?

Accordingly, I will conclude part two of this paper by proposing a Ramsian model of judging. Broadly speaking, a judge's vote contains information (independent of whatever reasons the judge may give to justify his or her vote), and judges should update their priors before casting their final and decisive votes, especially in close cases. (See generally Posner and Vermeule, 2016.) But is there any way of operationalizing these Bayesian insights? There is: "Ramsian voting" (or what I have called Bayesian voting; see Guerra-Pujol, 2019); that is, we would require each judge to not only state the reasons for his or her vote but also express his or her degree of belief in the vote.

In other words, A Ramsian system of judicial voting would replace the existing system of binary or "up or down" judicial voting with a new method of Bayesian voting in which judges would rate or score the strength of the legal arguments of the parties. With Ramsian voting, a judge would have to assign a numerical score reflecting his or her relative degree of belief or "credence" in what the proper outcome of an issue or case should be, depending on whether the judge is engaged in outcome-voting or issue-voting. [For an extended discussion of issue-voting versus outcome-voting by courts, see Post & Salop, 1992.]

To understand how Ramsian voting could work in practice, recall that one's degree of belief could be expressed in numerical terms anywhere in the range from 0 to 1:

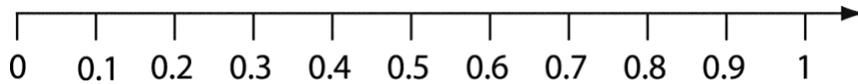

In summary, the higher the score, the greater the judge's credence or degree of belief. A score below 0.5, for example, would mean that the party with the burden of persuasion is not expected to prevail. A score above 0.5, by contrast, indicates that the party is expected to prevail, while a score of 0.5 means the judge is undecided about which party should prevail. Bayesian voting thus recognizes the subjective as well as the interdependent nature of law and legal interpretation.

This alternative method of voting goes by various names, including range voting (Smith, 2000), utilitarian voting (Hillinger, 2005), score voting (Lundh, n.d.), point voting (Hylland & Zeckhauser, 1980), and cardinal voting (Arrow, 1970), just to name a few variants. For my part, I have used the term "Bayesian voting" not only because law judicial decision-making in close cases is ultimately a subjective exercise in legal reasoning, but also to emphasize the close connection between this Ramsian method of judicial voting and the theory of "subjective probability" developed by such giants as Frank P. Ramsey and Bruno de Finetti. (See Galavotti, 1991.) Here, however, I will propose the term of art "Ramsian voting" in honor of Frank Ramsey.

*Conclusion*

I will conclude my review of Misak's biography with the following personal observations. Because of my own scholarly interests in Bayesian methods, my review of Misak's biography of Ramsey has been devoted mostly to Ramsey's contributions to probability theory. I now want to conclude my review with a confession and a conjecture. My confession is as follows: When I first read Misak's beautiful biography, I was at first sorely disappointed in two aspects of Ramsey's short life: his six-month sojourn in Vienna, and the open nature of his marriage. Let me explain.

With the benefit of 20/20 hindsight, i.e. knowing that Ramsey would have such a short time to live, I was personally upset by Ramsey's decision to squander no less than six months (!) of his short life to undergo psychoanalysis in Vienna. (For the record, I agree with Sir Karl Popper that psychoanalysis is "simply non-testable" and thus unfalsifiable pseudo-science. See Popper, 2002, p. 37.) Furthermore, I was also literally disheartened by the open nature of Ramsey's marriage to Lettice Baker because of my own normative or idealized view of romantic love (and my conflation of romantic love with marriage). Like a good Catholic, I believe marriage is a holy sacrament, or put in more secular terms, love should not be a matter of degree; true love requires sacrifice in order to signify what game theorists refer to as "credible commitment" (see, e.g., Schelling, 1980).

But now, having finished reading Misak's biography, I want to make a conjecture and perhaps even (like a good Bayesian!) update my priors regarding these two aspects of Ramsey's life. Although true love might be all or nothing, what if love in the real world is a matter of degree? (Or as an economist might put it, *what is the "optimal" level of love?*) Also, what if far from being a fruitless waste of time, what if it was Ramsey's extended exposure to psychoanalysis during his six-month sojourn in Vienna that somehow inspired him to develop his subjective approach to probability? After all, beliefs and desires--the raw materials, so to speak, of psychoanalysis--all play a critical role in Ramsey's subjective theory of probability. If so, his sojourn in Vienna was not a waste of time; it was a necessary precondition of his contributions to the world of probability theory!

Thank you, Cheryl Misak, for sharing your Frank Ramsey with us.

* * *

## Works Cited

Kenneth J. **Arrow**, *Social Choice and Individual Values* (2d ed. 1970).

Kenneth J. **Arrow**, *et al.*, "The Promise of Prediction Markets," *Science*, Vol. 320 (May 16, 2008), pp. 877-878.

Maya **Bar-Hillel**, "The base-rate fallacy in probability judgments." *Acta Psychologica*, Vol. 44, no. 3 (1980), pp. 211-233.

R. T. **Cox**, "Probability, frequency, and reasonable expectation," *American Journal of Physics*, Vol. 14, no. 1 (1946), pp. 1-13.

Bruno **de Finetti**, *Theory of Probability*, John Wiley & Sons (1974).

Maria Carla **Galavotti**, "The notion of subjective probability in the work of Ramsey and de Finetti," *Theoria*, Vol. 57 (1991), pp. 239-259.

F. E. **Guerra-Pujol**, "Chance and litigation." *The Boston University Public Interest Law Journal*, Vol. 21, no. 1 (2011), pp. 45-59.

\_\_\_\_\_\_\_\_\_\_\_, "Visualizing probabilistic proof." Washington University Jurisprudence Review, Vol. 7, no. 1 (2013), pp. 39-75.

\_\_\_\_\_\_\_\_\_\_\_, "Why don't juries try range voting," *Criminal Law Bulletin*, Vol. 51, No. 3 (2015), pp. 680-692.

\_\_\_\_\_\_\_\_\_\_\_, "The case for Bayesian judges," *The Journal of Legal Metrics*, Vol. 6, No. 1 (2019), pp. 13-20.

Alan **Hájek**, "Interpretations of Probability," in Edward N. Zalta, ed., *The Stanford Encyclopedia of Philosophy* (2019), available at https://plato.stanford.edu/archives/fall2019/entries/probability-interpret/.

Claude **Hillinger**, "The Case for Utilitarian Voting," Munich Discussion Paper, No. 2005-11 (2005), available at https://www.econstor.eu/bitstream/10419/104157/1/lmu-mdp_2005-11.pdf.

Oliver Wendell **Holmes**, Jr., *The Common Law*, Dover edition, (1991) [1881].

Aanund **Hylland** & Richard **Zeckhauser**, "A mechanism for selecting public goods when preferences must be elicited," Kennedy School of Government Discussion Paper 70D (1980).

John Maynard **Keynes**, *Treatise on Probability*, Macmillan (1921).

Patrick **Lundh**, "Score voting," The Center for Election Science, https://electology.org/score-voting (not dated).

Cheryl **Misak**, *Frank Ramsey: A Sheer Excess of Powers*, Oxford University Press (2020a).

\_\_\_\_\_\_\_\_\_\_\_, "Frank Ramsey: a genius by all tests of genius," History News Network (2020b), https://hnn.us/article/174250.

Karl **Popper**, *Conjectures and Refutations*, Routledge, 2nd edition (2002).

Eric A. **Posner** & Adrian **Vermeule**, "The votes of other judges," *Georgetown Law Journal*, Vol. 105, no. 1 (2016), pp. 159-190.

David **Post** & Steven C. **Salop**, "Rowing against *Tidewater*: a theory of voting by multijudge panels," *Georgetown Law Journal*, Vol. 80, no. 3 (1992), pp. 743-774.

Bertrand **Russell**, review of Keynes (1921), *The Mathematical Gazette*, Vol. 32, No. 300 (Jul., 1948) [1922], pp. 152-159.

Thomas **Schelling**, *The Strategy of the Conflict*, Harvard University Press, 2nd edition (1980).

Robert **Schlaifer**, *Probability and Statistics for Business Decisions*, McGraw-Hill (1959).

Warren D. **Smith**, "Range voting," unpublished manuscript (Nov. 28, 2000), available at https://www.rangevoting.org/WarrenSmithPages/homepage/rangevote.pdf.

Susan **Vineberg**, "Dutch book arguments," in Edward N. Zalta, ed., *The Stanford Encyclopedia of Philosophy* (2016), https://plato.stanford.edu/archives/spr2016/entries/dutch-book/.

Justin **Wolfers** and Eric **Zitzewitz**, "Prediction markets," *Journal of Economic Perspectives*, Vol. 18, No. 2 (Spring 2004), pp. 107-126.

* * *

# Appendix Re: Algebraic Notation

The Bayesian logic of Tables 2 and 3 and Figures 1 through 3 in the main text can be expressed using the following algebraic notation:

Let A denote a case decided at the pre-trial stage [striped]; let C denote a case decided at the trial or post-trial stage [dotted]; and let B denote a non-criminal government case.

$P(B, A) = P(A) \text{ times } P(B|A) = (.48) \text{ times } (.21) \cong .1$

$P(B, C) = P(C) \text{ times } P(B|C) = (.52) \text{ times } (.45) \cong .2$

The addition rule gives us the marginal probability as follows:

$P(B) = P(A) \text{ times } P(B|A) + P(C) \text{ times } P(B|C) = (.48) \text{ times } (.21) + (.52) \text{ times } (.45) \cong .1 + .2 \cong .3$

Lastly, the definition of conditional probability then gives us Bayes' Theorem:

$P(A|B) = P(A, B) \text{ divided by } P(B)$

and

$P(A, B) \text{ divided by } P(B) = [P(A) \text{ times } P(B|A)] \text{ divided by } [P(A) \text{ times } P(B|A) + P(C) \text{ times } P(B|C)]$

Next, recall that $P(A) = .48$; $P(B|A)] = .21$; $P(C) = .52$; and $P(B|C) = .45$. Accordingly, we can now plug in the relevant numerical values from our tables and figures in the main text as follows:

$P(A|B) = (.48 \text{ times } .21) \text{ divided by } [(.48 \text{ times } .21) + (.52 \text{ times } .45)]$

$P(A|B) = .1 \text{ divided by } (.1 + .2)$

$P(A|B) = .1 \text{ divided by } .3$

**$P(A|B) = .66$**

Note to this appendix: the numerical values above and in the main text can also be expressed more intuitively in discrete numbers, instead of fractions. For an example of such an approach, see Guerra-Pujol (2013).

<div align="center">* * *</div>

## Note to Law Review Editors

The Author will convert all scholarly citations and references in this book review into the Bluebook format upon acceptance of publication.